\documentclass[rmp,pre,floatfix,twocolumn]{revtex4}
\usepackage{graphicx}
\newcommand{\figwidth}{3.0 in}
\usepackage[greek,english]{babel}

\usepackage{babel}

\begin{document}
\latintext

\title{\bf Quantum formalism to describe binocular rivalry}

\author{Efstratios Manousakis\\
Department of Physics, Florida State University,\\
 Tallahassee, Florida, 32306-4350, USA and\\
 Department of Physics, University of Athens, \\
 Panepistimioupolis, Zografos, Athens, 157 84, Greece}

 
\begin{abstract}
On the basis of the general character and operation of the process
of perception, a formalism is sought to mathematically describe the
subjective or abstract/mental process of perception. It is shown that 
the formalism
of orthodox quantum theory of measurement, where the observer plays
a key role, is a broader mathematical foundation which can be adopted
to describe the dynamics of the subjective experience. The mathematical
formalism describes the psychophysical dynamics of the subjective
or cognitive experience as communicated to us by the subject. Subsequently,
the formalism is used to describe simple perception processes and,
in particular, to describe the probability distribution of dominance
duration obtained from the testimony of subjects experiencing binocular
rivalry. Using this theory and parameters based on known values of
neuronal oscillation frequencies and firing rates, the calculated
probability distribution of dominance duration of rival states in
binocular rivalry under various conditions is found to be in good
agreement with available experimental data. This theory naturally
explains an observed marked increase in dominance duration
in binocular rivalry upon periodic interruption of stimulus and yields
testable predictions for the distribution of perceptual alteration
in time. 

\end{abstract}


\maketitle

\section{Introduction}
\label{introduction} 

Several authors \cite{london,stapp,stapp1,stapp2,stapp3,nanopoulos,
penrose,foundations} including the founders of quantum
mechanics \cite{von-neumann,schrodinger,wigner,pauli}
have discussed the possible relation of consciousness to quantum theory.
It has also been argued that the mathematical formulation of quantum
mechanics is a broader foundation \cite{foundations} which can be
adopted to describe the most elementary mental events, i.e., the subjective
experience of the process of perception. In the present paper, on
the basis of the general character and operation of the process of
perception, it is suggested that the formalism of orthodox quantum
theory can be adopted to mathematically describe the subjective or
mental process of perception. We stress that the mathematical formalism
presented here does not aim at describing the brain dynamics 
of the observer as measured by an  observing instrument or by a second external 
observer observing
the brain of the first, but rather its aim is to describe the dynamics
of the subjective or mental experience as communicated by the first
observer himself. 
Namely, we seek a formulation to describe the dynamics of
the abstract or mental process of the subjective experience 
or the process of perception, for example,
the testimony of observers quantified by the recordings of a time
series of events occurring in their experience of binocular rivalry.
What is meant by these statements is clarified in the following section
by means of a simple example.

As described in
the following section, an attempt is made to give a precise mathematical
description of the character and operational nature of the process
of perception as experienced by subjects; it is argued and demonstrated
in Sec.~\ref{formulation} by means of examples that the mathematical
formalism of standard quantum mechanics, as we currently know it,
may be sufficient to quantitatively describe aspects of our conscious
experience and abstract mental processes. The difference between 
the earlier work on
the connection between quantum theory and consciousness and the present
work is that, here, we postulate and we present arguments to justify
it, that the formalism of quantum theory can be used to describe mathematically
the subjective or mental processes, such as the operation of perception
in binocular rivalry. The formalism is constructed with the goal to 
describe empirical data which are recordings of the experience of observers
to various stimuli, without a need to identify a material system where
the function of perception is manifested. Namely, the aim is to describe
the inner or mental experiences of observers, and, the goal is not
to describe an objectively existing physical system. In the present
paper, we explore further the quantitative connection of the formalism
of quantum theory using the formalism of standard quantum 
theory \cite{von-neumann,stapp,stapp1,stapp3,foundations}
and by applying the formulation to the well-known psycho-physical
phenomenon of binocular rivalry \cite{leopold,visualcompetetion,Tong06}. 
The theory presented here should find application in psycho-physical
phenomena where elementary aspects of the process of perception are
demonstrated. 

The problem of binocular rivalry has a long history,
and there are successful models \cite{freeman} to describe it using
concepts and tools of the classical computational neuroscience. The
aim of the present paper is not to show that the quantum formalism
is the only way to describe the experimental results on binocular
rivalry or that the above mentioned classical approach is insufficient.
Its purpose is, rather, to examine whether or not the available empirical
data pertaining to binocular rivalry fit well with the results of
a model where the formalism of quantum theory is applied to a simple
two state system. It is demonstrated that the application of
orthodox quantum theory works well and, therefore, might constitute
a useful foundation for understanding the connection of the central
nervous system to the subjective perception and qualia. It might,
therefore, be complementary to models of classical neuroscience. The
basic formalism used in the present paper is identical to that of
standard quantum theory in conjunction with the ideas and the interpretation
presented in the works of Von Neumann \cite{von-neumann} Stapp \cite{stapp1}
and Manousakis \cite{foundations}. However, the reader does not necessarily
have to resort to these works because we have made an attempt to make
this document self-contained. Namely, all the necessary mathematical
apparatus is introduced here in the following section and is adopted
to describe binocular rivalry.

\section{General Formulation}
\label{formulation}

\subsection{Potentiality and actuality}
\label{pc}

\begin{figure}[htp]
 \vskip 0.2 in 
\begin{center}
\includegraphics[width=\figwidth]{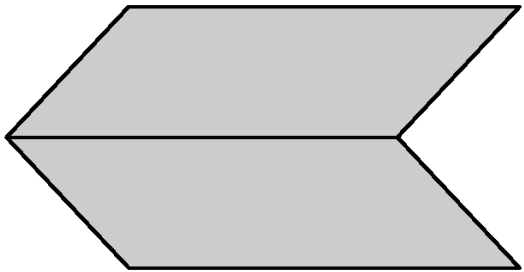}
\caption{\label{fig1}
An ambiguous  drawing: It  can be  interpreted as  a folded
paper with its edge inward or outward of the plane of the drawing.}
\end{center}
\vskip 0.2 in 
\end{figure}

In Fig.\ref{fig1} we present an ambiguous  drawing in order to provide a simple
example of the utility of the concept of {\it potential consciousness} which
will be  used in describing perception. Notice  that our consciousness
offers two potential interpretations  of this ambiguous figure: We can
perceive it as  a folded paper with its edge below  or above the plane
of the drawing. The reader is  encouraged (a) first to try to perceive
each of  these two interpretations separately, and  (b) while watching
the drawing to try to stick  to one of the two interpretations. Notice
that the suggestion (b) is difficult to follow because the  two different 
interpretations alternate  in perception every
few seconds.  Therefore, there are two different potential perceptions
of this figure, let us say, symbolically 1 and 2, the one in which the
edge  is below  and  the one  in which  the  edge is  above the  plane
respectively. If we  ask: ``what is the actual figure,  is one in which
the edge  is above or  below the plane?''  The answer to this question 
is:  ``neither'' or,
equivalently, the  answer can be also: ``both''. This reminds  us of the
question: ``where is  the electron in the atom?''  

One  of the claims of
the present  paper is that the best  way to describe the  state of our
consciousness between such perceptual events  can be written as: 
\begin{eqnarray}
| v \rangle = c_1 | 1 \rangle + c_2 | 2 \rangle = \left ( \begin{array}{c} 
c_1\\
c_2
 \end{array} \right ),
\label{state1}
\end{eqnarray}
where $|1 \rangle$ and $|2 \rangle$ denote two orthogonal unit 
vectors spanning a vector space and
the coefficients $c_1$, $c_2$, in  general, are complex numbers.  The reason for
using  this notation (known  in physics  as Dirac  notation) is  for a
notational convenience  which will become  clear in the  following few
steps. The  unit vector $| 1 \rangle$ corresponds to  one of the perceptions  
and the unit vector $| 2 \rangle$ to the  other  perception.   
The vector $| v \rangle$ given  by the  linear
combination in Eq.~\ref{state1}  describes the state of  potential 
consciousness which
corresponds to the stimulus which is  neither 1 nor 2; it becomes 
(or it is realized as) 1 or
2 through  the event of  a particular perception.  In a similar
sense the process of measurement in quantum theory (which, in the 
problem discussed here, corresponds
to the conscious realization or conscious event) brings into existence
a particular ``reality'' through the so-called wave-function ``collapse''.
We will see  in the
following  subsection  that the  coefficients $c_1$, $c_2$  are  
related to  the
likelihood  that  the  state 1  or  2  will  be  realized at  a  given
time.  Notice  that, as  special  cases  the  state described  by  
Eq.~\ref{state1}
includes  the ``pure'' states  1  and 2,  but  the  state of  potential
consciousness  is neither;  it corresponds  to the  linear combination
given  by  Eq.~\ref{state1}. This  vector $| v \rangle$, which  corresponds  
to the  state  of
potential  consciousness,  changes in  time  and  we  will discuss  the
equation which  gives its time evolution  in Sec.~\ref{evolve}.
When we  perceive the stimulus  as either 1  or 2, a  perceptual event
occurs which  brings into perception  one or the other  experience. We
may use the expression that,  when state (or interpretation) 1 occurs,
the state  of potential  consciousness $| v \rangle$ ``collapses'' 
to  $ | 1 \rangle$.  In  order to
describe the dynamics of such  perceptual or mental events we will use
the notion  of {\it potential  consciousness} and we  will show that  such a
tool is  very useful in  describing the evolution and  distribution of
dominance durations in binocular rivalry. In order to achieve that, we
need   an   equation   to   describe  the   time-evolution   of   this
state.  However, before  we discuss  this, we  should introduce  a few
other prerequisites.

Given  two vectors  
\begin{eqnarray}
| v \rangle = \left ( \begin{array}{c} 
c_1\\
c_2 \end{array} \right ), \hskip 0.3 in | v' \rangle = \left ( \begin{array}{c} 
c^{\prime}_1\\
c^{\prime}_2 \end{array} \right ), \nonumber
\end{eqnarray}
describing  two different  states of
potential  consciousness, we  define the  inner product  in  a Hilbert
space, in the familiar way, as  
\begin{eqnarray}
\langle v | v^{\prime} \rangle = (c^*_1,c^*_2)  \left ( \begin{array}{c} 
c^{\prime}_1\\
c^{\prime}_2 \end{array} \right )\equiv c^*_1 c'_1 + c^*_2 c'_2.
\label{norm}
\end{eqnarray}
Notice that in a Hilbert space,
the  complex conjugate of  the first  vector in  the inner  product is
used.  This allows  us to  define a  norm, a  measure, in  this space,
namely, the inner  product of a vector with itself  is a positive real
number  and  can  be used  as  the  measure  of  the ``length''  of  the
vector.  Also notice  in the  above notation  that we  may  regard the
vector $| v' \rangle$ as a $2\times 1$ matrix and we may regard the entity 
$\langle v | \equiv (c^*_1,c^*_2)$ as a $1\times 2$ matrix obtained
as the  transpose of $| v \rangle$, i.e.,  by converting its rows  
into columns and
taking the complex  conjugate of its elements. In  this way, the inner
product of two vectors is the matrix multiplication of a $1 \times 2$ 
with a $2 \times 1$ matrix
as shown in Eq.~\ref{norm} above. In addition, this justifies the origin of the
notation $\langle v | v^{\prime} \rangle$ used for  the inner product.  

Now, the  absolute magnitude of
the inner product between two vectors  can also be used to measure how
``close'' to  one another  these vectors are.  For example,  two vectors
having  zero  inner  product,  implies  that  these  two  vectors  are
orthogonal to each  other and this is the  largest possible difference
between two vectors.  Two vectors, each normalized to  unity (the norm
of each of  the vectors is 1), with inner product  equal to 1, implies
that these two  vectors are identical. The inner  product will be used
in  the  next  subsection where  we  need  to  compare two  states  of
potential consciousness.  

Therefore, in order to provide a description
for the process  of perception as {\it a mental} (as  opposed to a physical)
process we introduce a state of {\it potential consciousness} which includes
the  potentialities and  the  likelihoods for  all possible  conscious
experiences. This state evolves in  time and it drastically changes by
actual  conscious events.  To  describe this  mathematically, for  the
general  case, we  write the  state  of potential  consciousness as  a
vector or a linear superposition of all potential events $| i \rangle$, i.e.,
\begin{eqnarray}
| \psi \rangle = \sum_{i=1}^N
c_i | i \rangle  = \left ( \begin{array}{c} 
c_1\\
.\\
.\\
.\\
c_N
 \end{array} \right ),
\label{generalstate}
\end{eqnarray}
where $c_i$ are, in general, complex numbers and they are related to the
probability to  actualize the event $i$ whose  potentiality is represented
by the basis  vector $| i \rangle$. 
The set of all  possible potential events $| i \rangle$ forms
a complete basis in a Hilbert space describing all potential outcomes.

\subsection{Operation of Consciousness and observation}
\label{observation}

The previous example tells us, that  the stimuli have in some sense an
``unfathomable'' existence, or they  exist as potentialities, namely, we
can not talk  about them directly but only of  how they are perceived,
i.e., after they  have been {\it operated upon} by  the process of conscious
perception.   The cake  is not  sweet before  we experience  the taste
produced through the  reaction of our sensory apparatus;  before it is
tasted, it  is potentially sweet.  Hence, these potentialities  can be
actualized directly through the {\it action} of the corresponding ``organ'' of
perception  in  the  brain.  This operation  of  consciousness  called
perception or  measurement causes an  actual experience and  an actual
conscious  event.  

The  state  of potential  consciousness becomes  an
actuality, i.e.,  a conscious event  occurs, through the  operation of
conscious attention.  For example, imagine that you  are having dinner
with a  number of friends in  a round table  and several conversations
between sub-groups  take place simultaneously. When  you pay attention
or participate  in one  of the discussions,  while the  other physical
stimuli such as  voices come to your instruments  of perception, these
other  stimuli fade and  you become  aware only  of the  discussion to
which you  are paying conscious attention. One  can voluntarily switch
from one  conversation to another by  making an effort  to enhance the
probability of having  a ``collapse'' in a particular  desired sector of
potential  events.  

In  order to  describe this  operative  process of
consciousness we conceptualize consciousness (the thinker) as an actor
(or operator) which  acts on the state of  potential consciousness and
makes  actual   one  of  the  various   potentialities.  For  example,
consciousness can only perceive change, i.e., differences, variations,
alterations.  Imagine an observer who  had no prior experiences at all
and, suddenly,  the observer begins  to have a stream  of experiences.
The  observer  compares his  new  observation  with  a bank  of  prior
experiences  and this  makes observation  a comparative  process.  The
comparative character  of the  perceptual measurement process  is also
demonstrated  by   psychophysical  evidence  for   contextual  
effects \cite{Schwartz07}.
Namely, the perception of  a target
input depends strongly  on both its spatial context  (what surrounds a
given object) and its temporal  context (what has been observed in the
recent past).  At first, this  idea that we only observe change, might
be difficult to  accept because we know that we are  able to observe a
{\it static}  object. In  actuality, however,  we  are able  to observe  the
object  because of the  constant change  caused on  our retina  by the
light  coming  from the  object.   This  change  produces an  electric
potential difference which  will cause a neuron to  fire.  In order to
be able to  see the same static object  ``continuously'', the neuron has
to   be   charged   anew   and   when   it  ``fires'',   it   causes   a
perception.  Therefore,   one  of  the  functions   or  operations  of
consciousness is to  be able to perceive a  change. The operator which
measures  change in  space  is the  differential  operator $\nabla$, 
while  the
operator which measures temporal changes  is $\partial_t$ 
and both act on the state
of potential consciousness $| v \rangle$.  

However, we need to answer the following
fundamental question:  how does consciousness see or  measure a change
in the absence of an a  priori content of memory to be used for comparison? What
is the  standard, or  the measure, which  consciousness should  use in
order  to measure  the new  state? Namely,  consciousness  operates an
inquiry (a question) which is  represented by a particular operator 
$\hat Q$ (a
matrix) which acts on the  state of potential consciousness $|v \rangle$
and during
this  operation a  new,  transient, state  of potential  consciousness
results which  we represent as $| v^{\prime} \rangle =
\hat Q | v \rangle$ . Only one  state from the set  of the
various potentialities will be actualized after the measurement; which
one?  

Let us analyze this question in terms of the example given above
regarding the ambiguous figure.  In this case, consciousness poses the
question: Q  = ``Is the  observed in state $| 1 \rangle$ ?'' 
($| 1 \rangle$ and $| 2 \rangle$ represent  the two
alternative states corresponding  to the alternative percepts observed
when observing  the folded paper in Fig.1).  The  projection
operator associated with the above inquiry Q is 
\begin{eqnarray}
\hat Q =  \left ( \begin{array}{cc} 
1 & 0\\
0 & 0 \end{array} \right ),
\label{inquiry}
\end{eqnarray} 
i.e., depending  on the state, it acts as follows:
\begin{eqnarray}
(a)& & \mathrm{if} \hskip 0.1 in | v \rangle = | 1 \rangle =  \left ( \begin{array}{c} 
1\\
0 \end{array} \right ), \hskip 0.1 in \mathrm{then}, \hskip 0.1 in \hat Q | 1 \rangle = | 1 \rangle, \nonumber \\
(b)& & \mathrm{if} \hskip 0.1 in | v \rangle = | 2 \rangle =  \left ( \begin{array}{c} 
0\\
1 \end{array} \right ), \hskip 0.1 in \mathrm{then}, \hskip 0.1 in \hat Q | 2 \rangle = 0. \nonumber
\end{eqnarray}
Now,  how does  consciousness
measures the outcome of this inquiry? It has no measure or standard by
which to carry out the  measurement, other than the state of potential
consciousness just before  the inquiry, i.e., $| v \rangle $ given 
by  Eq.~\ref{norm}. Namely,
the likelihood  that the  answer to this  question is ``yes''  should be
given   by   the   comparison    of   the   state $| v^{\prime} \rangle=
\hat Q | v \rangle$  with   the   state $| v \rangle$
itself. Mathematically, this is evaluated by the inner product between
theses two potential states,  i.e., $\langle v | v^{\prime} \rangle$ 
 and this
notation is defined  by Eq.~\ref{norm}. As discussed in the previous 
subsection the inner product measures how close to each other two
states of potential consciousness are. Thus, during  the mental event or 
experience or ``collapse'' of the state of potential consciousness, 
the expected value of an observable $\hat Q$ is 
\begin{eqnarray}
\langle \hat Q \rangle = \langle v | \hat Q | v \rangle.
\label{answer}
\end{eqnarray}
Namely, this is the probability for the
answer to the above inquiry to be ``yes''.

\subsection{Evolution of potential consciousness}
\label{evolve}

When we  consider the  time evolution  of any vector $|\psi(t)\rangle$
 it is  useful to define  a  different   set  of  vectors  forming 
a complete basis of the same Hilbert
space. Such a basis set is formed by all the
 periodic  states $| \nu \rangle_t$ characterized by
frequency $\omega_{\nu}$ (and
period $T_{\nu} = 2\pi/\omega_{\nu}$),  i.e., 
$| \nu \rangle_{t+T_{\nu}} = | \nu \rangle_t$; the  time
evolution of these states is given by
\begin{eqnarray}
| \nu\rangle_t = e^{\pm i \omega_{\nu} t} | \nu\rangle; \hskip 0.3 in | \nu \rangle = | \nu \rangle_0,
\end{eqnarray}
and we define an operator $\hat \omega$  which is a diagonal  matrix in the
basis $| \nu \rangle$ with eigenvalues $\omega_{\nu}$, i.e.,
\begin{eqnarray}
\hat \omega | \nu \rangle = \omega_{\nu} | \nu \rangle.
\end{eqnarray}
The  matrix $\hat \omega$  is  Hermitian, namely, 
$\langle i | \hat \omega | j \rangle = (\langle j | 
\hat \omega | i \rangle)^*$ because  it has  real
eigenvalues. The  time-dependent state of  potential consciousness can
be expanded in a Fourier transformation as follows:
\begin{eqnarray}
| \psi(t) \rangle = \sum_{\nu} e^{i \omega_{\nu} t} c_{\nu} | \nu\rangle,
\label{vmotion}
\end{eqnarray}
where $c_{\nu}$ are Fourier expansion coefficients and,  here, the sum is
over both negative and positive $\omega_{\nu}$.  

Using the definition of $\hat \omega$, Eq.~\ref{vmotion} can
be written  as follows:
\begin{eqnarray}
| \psi(t) \rangle = e^{i \hat \omega t} \sum_{\nu}  c_{\nu} | \nu\rangle,
\end{eqnarray}
 which  implies that
\begin{eqnarray}
| \psi(t) \rangle = e^{i \hat \omega t} | \psi(0) \rangle,
\label{vmotion2}
\end{eqnarray}
 and  this is
equivalent  to  the following  evolution  equation:  
\begin{eqnarray}
\hat{\omega}|\psi_{t}\rangle=i\partial_{t}|\psi_{t}\rangle.
\label{motion}\end{eqnarray}
We  have,
therefore, transformed our original problem  into one in which we seek
an  operator $\hat \omega$ for  the particular  perception problem  
at hand.  In the
example discussed previously $\hat \omega$ is  a general 
$2 \times 2$  Hermitian matrix.  

This is a  Schr\"odinger-like equation of  motion describing 
the  dynamics of
cognitive  processes.  Hence,  the  state of  potential  consciousness
evolves  in a  similar way  as the  state vector  evolves  in standard
quantum mechanics \cite{von-neumann},  namely, as given by Eq.~\ref{vmotion2} 
where
the  time  displacement (or  evolution  operator) $\hat U$  is  related 
to  the frequency operator $\hat \omega = \hat H/\hbar$ 
(where $\hat H$ is  the Hamiltonian in quantum mechanics and $\hbar$
is  Planck's constant)  as follows 
\begin{eqnarray}
\hat U = e^{i \int_0^t \hat \omega(t') dt'},
\end{eqnarray}
namely,  the  frequency operator
(which in general  can change as a function of  time) is the generator
of  infinitesimal  translations in  time.  In  our approach,  Planck's
constant does not  enter in our equation of  motion.  

Conscious events
that occur  are identified with the quantum  mechanical ``collapses'' of
the wave function, as specified by the orthodox quantum theory and the
meaning  of this concept  was explained  in subsection \ref{pc} above.  
The wave
function,  between  such  perceptual  events,  describes  a  state  of
potential   consciousness   which    evolves   via   the   Schr\"odinger
equation. When  a perceptual event is observed,  where a wave-function
``collapse'' occurs  through the  process of measurement,  it actualizes
the corresponding neural correlate of consciousness (NCC) brain state.

\section{Application to Binocular Rivalry}
\label{rivalry} 

\subsection{Time evolution of potential consciousness}
\label{evolve:BR}

In binocular rivalry \cite{visualcompetetion,leopold,visualawareness},  
which is an example  of multi-stable
perception \cite{orbach},  two different images  are presented
dichoptically to awareness.  These two  images are the stimuli for two
different potential percepts and form  the basis vectors in terms of which 
to express
the  general   state  of   potential  consciousness.  The   action  of
consciousness, i.e.,  through conscious  attention, at a  higher level
causes the perception of a  particular potentiality as a real event in
consciousness  by  activating  one  particular percept.  Through  this
operation of consciousness, the conscious percept (or symbol) acts and
activates the  corresponding neural correlates  of consciousness (NCC)
in the nervous system. Furthermore, a sequence of ordered events arise
in consciousness as a time series $t_1,t_2,...,t_n$ , at each one 
of which a perceptual
change occurs and this defines  a flow in consciousness.

{\it State vector formulation:} 
For clarity, we begin with the state vector formulation and
below, the problem is also discussed using the density matrix formalism
for completeness. We imagine two potential states of
consciousness denoted as $|1\rangle$ and $|2\rangle$ which correspond
to the two states which can be realized in the case of binocular rivalry,
which are associated with two distinct NCC brain states. 
An arbitrary state of potential
consciousness will be written as in Eq.~\ref{state1} with
\begin{eqnarray}
|c_1|^{2}+|c_2|^{2}=1,
\end{eqnarray}
 such that, after the process of projection, the probability for either
distinct state to become actual,
is unity. Therefore, in binocular rivalry, we have two rival states
of consciousness associated with percepts $|1\rangle$ and
$|2\rangle$ which, in turn, correspond to two different NCC states. 

For binocular rivalry, and more generally for a two-state problem
$\hat{\omega}$ is a $2\times2$ matrix, \begin{eqnarray}
\hat{\omega}=\left(\begin{array}{cc}
\langle1|\hat{\omega}|1\rangle & \langle1|\hat{\omega}|2\rangle\\
\langle2|\hat{\omega}|1\rangle & \langle2|\hat{\omega}|2\rangle\end{array}\right)=\left(\begin{array}{cc}
\epsilon_{1} & h_{12}\\
h_{21} & \epsilon_{2}\end{array}\right).\label{frequencyoperator}\end{eqnarray}
 As discussed in Sec.~\ref{evolve}, the operator $\hat{\omega}$
must be a Hermitian operator, therefore, $h_{21}=h_{12}^{*}$. 
To simplify the problem, let us consider, the case where 
$\epsilon_{1}=\epsilon_{2}=\epsilon$
(namely, the two rival states are ``degenerate'', i.e., there is
no preference for or bias against one or the other).

In order to solve Eq.~\ref{motion} in the case where $\hat \omega$ is 
time-independent, we generally proceed \cite{von-neumann,bohm} by
finding the eigenvalues $\omega_{\nu}$ and the eigenstates $| \nu \rangle$
of the operator $\hat \omega$ and by expanding the initial state
$| \psi(0)\rangle$ in the basis formed by the eigenstates, i.e.,
\begin{eqnarray}
| \psi(0) \rangle = \sum_{\nu} c_{\nu} | \nu \rangle,
\end{eqnarray}
and the general time-dependent solution is given by Eq.~\ref{vmotion}.

Let us suppose that the initial state is just one of the two states,
for example state $|1\rangle$. As long as there is no observation
(conscious or involuntary attempts to project), the state of potential
consciousness as a function of time 
is in a state given as follows: \begin{equation}
|\psi_{t}\rangle=\cos(\bar{\omega}t)|1\rangle-ie^{-i\phi}\sin(\bar{\omega}t)|2\rangle.
\label{potential}\end{equation}
 The frequency $\bar{\omega}=|h_{12}|$ and it is given as $\bar{\omega}=2\pi/T$;
$T$ is the characteristic period of the instrument of consciousness.
The phase $\phi$ characterizes the off-diagonal matrix elements $h_{12}=\bar{\omega}e^{i\phi}$.
The diagonal elements $\epsilon$ of $\hat{\omega}$, in the case
where $\epsilon_{1}=\epsilon_{2}=\epsilon$, only contribute to an
overall phase which has no observable effects.

Let us consider the inquiry Q: ``is the observed in state
$|1\rangle$?'', which, for the binocular rivalry case, is represented 
by the same operator $\hat Q$ given by Eq.~\ref{inquiry}. 
The expected answer to the inquiry Q
is obtained by Eq.~\ref{answer}, and by straightforward substitution of
Eq.~\ref{potential} for the state $| \nu \rangle$ we find that: 
\begin{eqnarray}
\langle\hat{Q}\rangle=\cos^{2}(\bar{\omega}t).
\label{average}
\end{eqnarray}

{\it Density matrix formulation:} Now, let us turn our 
discussion to the formulation of the problem
using the density matrix formalism. The density matrix for a general
two state system is given by \begin{eqnarray}
\hat{\rho}=\left(\begin{array}{cc}
\langle1|\hat{\rho}|1\rangle & \langle1|\hat{\rho}|2\rangle\\
\langle2|\hat{\rho}|1\rangle & \langle2|\hat{\rho}|2\rangle\end{array}\right)=\left(\begin{array}{cc}
\rho_{11} & \rho_{12}\\
\rho_{21} & \rho_{22}\end{array}\right).\label{densitymatrix}\end{eqnarray}
 The equation of motion satisfied by the density matrix operator is
\begin{eqnarray}
i\partial_{t}\hat{\rho}=[\hat{\omega},\hat{\rho}],\label{densitymatrixequationofmotion}\end{eqnarray}
 where $\hat{\omega}$ is the frequency operator ($\hat{\omega}=\hat{H}/\hbar$)
given by Eq.~\ref{frequencyoperator} for the binocular rivalry case.
A measurement or observation takes place by means of a conscious inquiry
which is represented by a projection operator $\hat{Q}$ (i.e., $\hat{Q}^{2}=\hat{Q}$).
For our case of the binocular rivalry the projection operator is given
by Eq.~\ref{inquiry}. The result of a measurement is given as \begin{eqnarray}
\langle\hat{Q}\rangle=\frac{{Tr(\hat{\rho}\hat{Q})}}{{Tr(\hat{\rho})}}.\label{trace}\end{eqnarray}
 If after such a measurement occurs and the answer is {}``yes''
the state of potential consciousness is represented by a density matrix
which equals the above projection operator $\hat{Q}$. If after the
measurement is ``no'' the state of potential consciousness is
represented by a density matrix given by $\hat{1}-\hat{Q}$.

Let us assume that at time $t=0$ the state of potential consciousness
corresponds to the state $|1\rangle$, i.e., to a density matrix $\hat{\rho}=\hat{Q}$.
It is straightforward to show that the density matrix $\hat{\rho}(t)$,
which satisfies the equation of motion (\ref{densitymatrixequationofmotion})
for any time $t$, is given by \begin{eqnarray}
\hat{\rho}=\left(\begin{array}{cc}
\cos^{2}(\bar{\omega}t) & -i\sin(\bar{\omega}t)cos(\bar{\omega}t)\\
i\sin(\bar{\omega}t)cos(\bar{\omega}t) & \sin^{2}(\bar{\omega}t)\end{array}\right).
\label{densitymatrixsolution}\end{eqnarray}
By substituting this expression for the density matrix  
in Eq.~\ref{trace} and using the expression given by Eq.~\ref{inquiry}
for the operator $\hat Q$, we obtain the same
result for $\langle \hat Q \rangle$ 
as in Eq.~\ref{average} above.

\subsection{Measurement/projection in Binocular Rivalry}

\label{measure}

In order for an event to enter the stream of consciousness, a projection/measurement
(see Sec.~\ref{observation} in the introduction of this paper) 
must occur. Let us
now assume that after time $\delta t$ an observation or measurement
is carried out where the state of consciousness is observed by operating
(operationally asking) the question Q: ``is the observed
in state $|1\rangle$?''

This is asked by applying the operator $\hat{Q}=|1\rangle\langle1|$
(Eq.~\ref{inquiry})
on the state of potential consciousness $|\psi_{t}\rangle$ and the
probability to observe the state $|1\rangle$ is given by $\langle\psi_{t}|\hat{Q}|\psi_{t}\rangle=\cos^{2}(\bar{\omega}\delta t)$ (See Eq.~\ref{average}).
The operation of the inquiry on the state of potential consciousness
actualizes the state $|1\rangle$ (or $|2\rangle$) with probability
$\cos^{2}(\bar{\omega}\delta t)$ (or $\sin^{2}(\bar{\omega}\delta t/T)$).
If $\delta t<<T/4$ there is high probability to observe the state
$|1\rangle$. Therefore, if we keep making conscious or involuntary
attempts to observe frequently, the same state is projected (the quantum
Zeno effect).

\begin{figure}[htp]
 \vskip 0.2 in 

\begin{center}
\includegraphics[width=\figwidth]{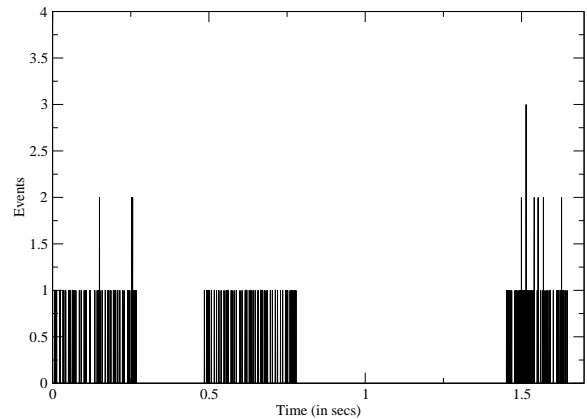}
\end{center}
\caption{\label{burstillustrate} Illustration of the neuronal activity or
measurement events needed to understand the distribution of dominance
duration in binocular rivalry. Notice that there are the following
time scales. (a) A long time scale $T_{b}$, the average time between
neuronal firing bursts, (b) the duration of the burst $\delta T$,
and (c) the average firing rate $r_{f}=1/T_{f}$ within each burst.
Notice that $T_{b}$ is of the order of 1 sec, $\delta T$ is a fraction
of a second and $f_{s}$ of the order of $100Hz$. \cite{singer1,gray,singer2,singer3}}

\vskip 0.2 in 
\end{figure}

We will assume that in an interval $0<t<t_{max}$ a series of observing
events occur each one of which causes the neurons to fire in bursts
as illustrated in Fig.~\ref{burstillustrate}. Let us assume that
the observations take place at time instants $t_{1},t_{2},...,t_{n},...,$
and these instances have been picked from a given distribution, an example
of which is given in Fig.~\ref{burstillustrate}. 
The probability that the initial state during the first $i-1$ observations
at the instances $t_{1},t_{2},...,t_{i-1}$ was repeatedly found to
be $|1\rangle$ and found to switch to $|2\rangle$ at time $t_{i}$,
is 
\begin{eqnarray}
w_{i}=\sin^{2}(\bar{\omega}(t_{i}-t_{i-1}))\prod_{j=1}^{i-1}
\cos^{2}(\bar{\omega}(t_{j}-t_{j-1}))).
\end{eqnarray}
In this case the dominance duration of the initial state is $t_{i}$.
Therefore, the probability density of dominance duration $t$ is given
as $P(t)=\sum_{i=1}^{n}w_{i}\delta(t-t_{i})$. This probability density
can be calculated using a random walk Monte Carlo process where the
time instants of the measurement events $t_{1}$, $t_{2}$,...,$t_{n}$,
are chosen from a desired distribution similar to that of Fig.~\ref{burstillustrate}.
One simple choice of such a sampling technique is to select the first
$N_{f}$ events which constitute the first burst as $t_{i}=t_{i-1}+\delta t$
(with $t_{0}=0$) where the variable $\delta t$ is chosen from a
Gaussian distribution with variance $T_{f}$ (or as a random variable
uniformly distributed in the interval $[0,T_{f}]$). Similarly, the
beginning of the next burst is chosen from a Gaussian distribution
with variance $T_{b}$ (or as a uniform random variable in the interval
$[0,T_{b}]$). A sample train of three such bursts produced using
Gaussian distributions with $T_{f}=5msec$, $N_{f}=60$ and $T_{b}=0.9sec$
is shown in Fig.~\ref{burstillustrate}.

In Section~\ref{appendixa}(Appendix) we present a detailed study
of this model and how the results differ by changing the neuron firing
rates.

\section{Results and comparison with experiment}
\label{results}

\subsection{Probability distribution of dominance duration}



%

%
\begin{figure}[htp]
 \vskip 0.2 in 

\begin{centering}
\includegraphics[width=\figwidth]{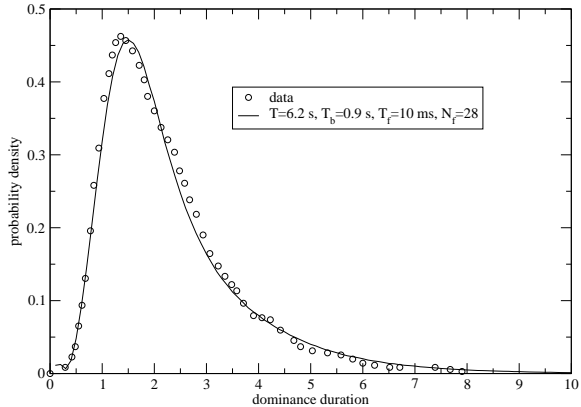}
\par\end{centering}

\caption{\label{data} Comparison with the data from Lehky is made using $T=6.2sec$,
$T_{b}=0.9sec$ and $T_{f}=10msec$ and $N_{f}=28$ (only the combination
$\delta T=T_{f}N_{f}=0.28sec$ is relevant). }
\vskip 0.2 in 
\end{figure}

\begin{figure}[htp]
 \vskip 0.2 in 
\begin{center}
\includegraphics[width=\figwidth]{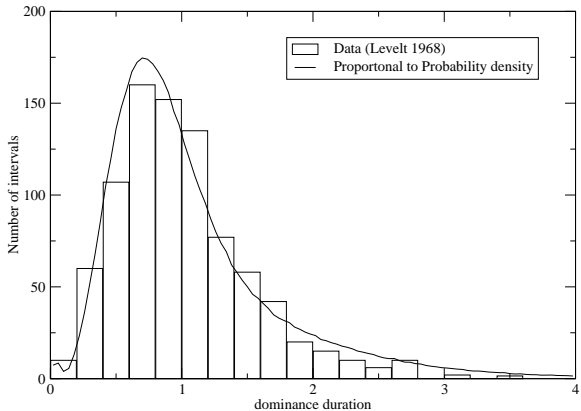}
\caption{\label{levelt} Comparison with the data from Levelt using $T_{b}=0.5$,
$\delta T=0.1$, and $T=3$. Time in Ref.~\cite{levelt} is in units
of the mean dominance duration. The time scale $T$ is the period
characterizing the evolution of the state of potential consciousness
(See Section~\ref{evolve}).}
\end{center}

\vskip 0.2 in 
\end{figure}

\begin{figure}[htp]

\begin{centering}
\vskip 0.2 in \includegraphics[width=\figwidth]{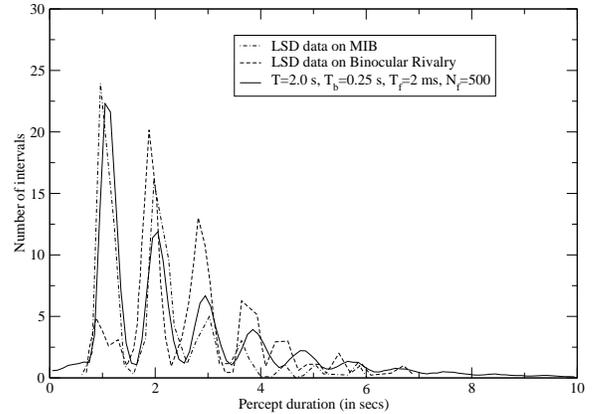} 
\par\end{centering}

\caption{\label{figurelsd} Comparison of the theoretical distribution of dominance
duration (solid line) with the data from subjects under the influence
of the hallucinogen LSD. The theoretical curve is obtained using $T=2.0sec$,
$T_{b}=0.25sec$ and $\delta T=T_{f}N_{f}=1sec$.}

\begin{centering}
\vskip 0.2 in 
\par\end{centering}
\end{figure}

In Fig.~\ref{data}, the distribution of dominance duration obtained
using the present model is compared with the data obtained by
Lehky \cite{rivalrydata} and in Fig.~\ref{levelt} it is compared 
with the data obtained by
 Levelt \cite{levelt}. The values of the parameters used in the
fit, which are given in the figure captions, are reasonable. Neuron
firing frequencies of the order of $100Hz$ have been used in these
calculations. As discussed in the appendix the probability distribution
of dominance duration is not sensitive to the precise value of the
spike frequency as long as the burst duration is kept constant. Namely,
by varying the spike frequency by even a factor of two or more (within
the observed range of neuron firing frequency for the mammalian brain \cite{fries})
and keeping the burst duration constant, we obtain the same probability
distribution of dominance duration. The value of the frequency $T$
is determined by setting the scale of time for the graph; the other
parameters, which define the neuron firing pattern, are the average
time between bursts $T_{b}$ and the burst duration $\delta T$ and
their values are consistent with the neuron firing pattern observed
for the alert cat \cite{singer1,singer2,singer3}. The firing pattern
of single cells in the striate cortex of macaque monkeys \cite{monkey}
indicate similar neuron firing patterns. The firing patterns might
be somewhat different for neurons of the human cerebral or extra striate
cortex, which is the case of our interest; therefore, the values found,
in order to achieve good agreement with the observed PDDD, are reasonable.

The value of the oscillation period $T$ is of similar magnitude to
a value known as a characteristic time scale for a low-frequency mechanism
that binds successive events into perceptual units \cite{poppel,shortmemory}.
In a study of the human auditory sensory system by examining the mismatch
negativity (MMN) using SQUID magnetometry \cite{neurosquid} it was
found that the amplitude of the MMN as a function of the inter-stimulus
interval becomes large at a time scale of the order of $3sec$. More
recently direct functional MRI measurements \cite{fmri,fmri2,fmri3,fmri4}
demonstrate that the perceived alteration is accompanied by responses
in extra-striate cortex areas which are characterized by such time
scales. In addition, oscillations at such very low frequencies which
correspond to a fraction of a $Hz$ have been observed in EEG patterns
during the deep stages of sleep of humans \cite{eeg0,eeg1,eeg2} and
of the cat cerebral cortex \cite{eegcat}. 


In studies \cite{lsd} with a subject who took the hallucinogenic drug LSD
an oscillatory behavior of the distribution of dominance duration
was revealed. In Fig.~\ref{figurelsd}, the distribution obtained
for the case of binocular rivalry and motion-induced blindness (MIB)
from the subject under the influence of LSD is compared with the present
model when we use the values of the parameters shown in the inset.
Therefore, we conclude that, within the context of our  model,
the hallucinogenic drug changes the burst frequency and the characteristic
period $T$ significantly, which places the observing conscious apparatus
into a regime where oscillations are transparent. More recent 
studies on the influence of hallucinogenic 
drugs \cite{carter1,carter2} indicate that it might be
in the ``rebound-phase'' (a phase reached several hours after
the peak of the drug influence) that a state of the type indicated
in Fig.~\ref{figurelsd} and in Ref.~\cite{lsd} may be realized
and that this state may not be related at all to the state of consciousness
reached at the peak of the LSD influence. More specifically, 
in Ref.~\cite{carter1},
it was found that the response of subjects at the peak of the drug
influence was found to be significantly slower; however, switches
for all subjects became increasingly faster, such that six hours later
some subjects were switching at intervals that were shorter, more
regular and more rhythmic than their pretest levels. Therefore, this is also
consistent with the results of Ref.~\cite{lsd} where the subject
was tested many hour later than the peak of the drug influence.

We would like to comment about the so-called weak  quantum 
theory (WQT)  \cite{atmanspacher1}  which  is an
attempt to generalize the formal framework of quantum theory in such a
way that complementarity and entanglement might be useful in a broader
context. The WQT has been used  to describe ambiguous 
perception \cite{atmanspacher2,atmanspachernew}. There are the following
important   differences  between   the  present   work  and 
WQT  and its applications \cite{atmanspacher2,atmanspachernew}: (a)  
The formalism used in the present paper to
describe the dynamics of the  mental processes, is identical to that of
orthodox quantum theory and we found  no need to resort to some
other mathematical  foundation in the present paper.  In addition, there
are important differences between the two mathematical models, as applied 
to the problem of multi-stable perception, namely between the application
of WQT  and our work.  (b) In  the present  paper the
problem of binocular  rivalry is studied rather than  the problem of
ambiguous perception.

\subsection{Periodic removal of stimulus}
\label{removal}
\begin{figure}[htp]
 \vskip 0.2 in 

\begin{centering}
\includegraphics[width=\figwidth]{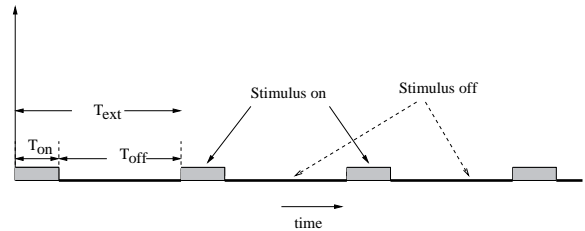} 
\par\end{centering}

\caption{\label{interval} The periodic switching on and off of the rival stimuli.
The rival figures are dichoptically presented to awareness for an
interval $T_{on}$ and they are removed for a blank interval $T_{off}=T_{ext}-T_{on}$
where $T_{ext}$ is the period of the externally modulated stimulus.
Random observations take place during the on-intervals with an average
frequency $f_{s}=1/T_{s}$.}

\vskip 0.2 in 
\end{figure}

Next, we discuss the results
of Ref.~\cite{stablerivalry} where the stimulus was periodically
removed as shown in Fig.~\ref{interval}. Namely, the stimulus appears
and disappears periodically and both rival images are presented to
awareness for a time interval $T_{on}$, and for a time interval $T_{off}$
both images are removed. It was found that the frequency of perceptual
alterations can be slowed down, and even brought to almost a standstill.
These experimental results are very important because they provide
clear support for the present theory. 



Here, we are dealing with a frequency operation and a time-evolution
operator which changes with time as measured by an external clock.
Namely, the frequency operator is given by a constant operator 
$\hat{\omega}(t)=\hat{\omega}$ (which is a $2\times 2$ matrix),
in the on-stimulus time intervals and  $\hat{\omega}(t)=0$
in the off-stimulus intervals. When the rival images are presented,
the evolution occurs according to the frequency operator $\hat{\omega}$
given by Eq.~\ref{frequencyoperator} and when they are totally absent
the frequency operator is a constant $C$ (which can be taken to be
zero). The general solution of the Schr\"odinger-like equation, when
$\hat \omega(t)$ is time dependent, is given by 
$|\psi_{t}\rangle=e^{i\int_{0}^{t}dt'\hat{\omega}(t')}|\psi_{0}\rangle$,
which implies that $|\psi(t)\rangle$ in the on-stimulus intervals
evolves as before and during the off-stimulus interval the state of
potential consciousness {}``freezes''.

\begin{figure}[htp]
 \vskip 0.2 in 

\begin{centering}
\includegraphics[width=\figwidth]{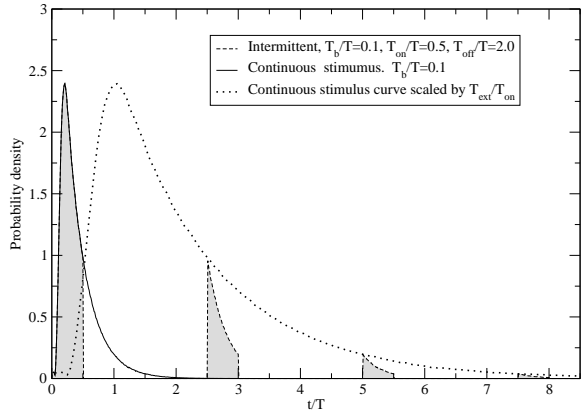} 
\par\end{centering}

\caption{\label{intermittent} Probability distribution of dominance duration
in binocular rivalry with intermittent stimulus. The average time
between bursts $T_{b}/T$ is taken to be $0.1$ and the width of the
burst is $\delta t/T=0.05$. (a) Solid line: Continuous stimulus.
(b) Gray-shaded: Intermittent presentation of stimulus. (c) Doted-line:
The distribution of part (a) where the time axis is scaled by a factor
$(T_{on}+T_{off})/T_{on}$. }

\vskip 0.2 in 
\end{figure}

In Fig.~\ref{intermittent} the probability distribution of the dominance
duration is shown. Notice that the probability distribution which
corresponds to continuous presentation of the stimulus (solid line),
splits into separate parts (gray-shaded) with duration $T_{on}$,
when the stimulus is presented intermittently. However, if we eliminate
these blank intervals (as if they did not exist) and the separate
parts are placed side by side, the result is the same as the distribution
obtained with continuous presentation of the stimulus. This result
has a number of consequences and we would like to mention the following
two: (a) For fixed $T_{on}$, the average dominance duration grows
linearly with the blank duration $T_{off}$. (b) For fixed blank duration
$T_{off}$, the average dominance duration decreases with increasing
$T_{on}$. These conclusions agree well with the available experimental
results \cite{orbach,stablerivalry}. %

We would like to note that when the stimulus turns on after the blank
duration, the change itself is expected to cause a measurement. Our
simulation shows that, if at the beginning of the {}``stimulus on''
periods we always start with a burst of $N_{f}$ measurements and
the bursts which follow this initial burst are selected according
to the same probability distribution, there is additional broadening
of the distribution, thus, further increasing the dominance duration.
In this case the above mentioned features (a) and (b)
are only approximately valid.

\section{Discussion}

The present explanation of the phenomenon observed, when the stimulus
was periodically removed, is based on the fact that our theory places
the attention of consciousness higher, in the hierarchy of consciousness,
than the two stimulated neural correlates in the brain. 
If we interrupt the external process which presents to awareness the
mixture of the two potential states, the state of potential consciousness
remains (as memory, with no oscillatory evolution) on the most recently
collapsed percept; namely, because of the interruption of the external
stimulation of the rival state, there is no reason for alterations
of the state of potential consciousness because there is no likelihood
associated with the other percept. Therefore, when a new stimulation
occurs, where both conscious symbols are presented to awareness, evolution
of the state of potential consciousness begins starting from the previously
collapsed state. Thus, shortly afterwards, the state of potential
consciousness will collapse again in the previously collapsed state
with high probability. Hence, when the external presentation to awareness
of the stimulus is halted, the time evolution of the state of potential
consciousness {\it stops}, and the {\it perceived time} stands
still.

{\it Role of Attention:} There are recent studies \cite{ooi,mitchell,chongblake}
where attention can bias the initial selection of state in binocular
rivalry toward the attended state (or stimulus). In our formulation
attention is a fundamental aspect of consciousness. Attention is the
voluntary or involuntary preparation of the state of potential consciousness
with increased likelihood, which causes biased preference, for a certain
event or events to occur. If the initial state of potential consciousness
is prepared in such a way to be the state $|1\rangle$, this state
will be a preferred state when an event occurs soon afterwards.

There are two types of factors which can enhance {\it attention}.
1) {\it Bottom-up factors}: If we make the diagonal terms of the
operator $\hat{\omega}$ different, the probability distribution of
dominance durations for state $|1\rangle$ and for $|2\rangle$ would
be equal. However, the transition rates would be smaller, a fact that
prolongs dominance durations (because there is a potential barrier
to be crossed for the oscillations to occur). This is in agreement
with the experimental conclusions \cite{chong,meng}. 2) {\it Top-down
factors}: Instructing the observer to pay attention to one particular
perceptual state influences and modulates the frequency of measurements.
When one eye's stimulus is strengthened (e.g. by increasing contrast),
the mean dominance duration of the other (unaffected) eye decreases \cite{levelt}.
In this case the rate of observations $f_{s}$ is different for each
of the two eyes. The contrast difference causes the {}``observer''
to increase the rate of observations of one of the states when that
state is being experienced.

\section{The issue of decoherence}

If we consider the formalism of quantum theory as a mathematical
foundation that describes the microscopic world, it is very
difficult to make a convincing argument that the
brain dynamics is governed by quantum theory because
of the issue of decoherence \cite{Zeh,Joos,Giulini,tegmark}.
In this work, however, we argued that quantum theory is a 
broader foundation and it can describe the subjective experience of
our thought process and more generally,  the perception process. 
We have reasoned that we can use the formalism of 
quantum mechanics to describe the testimony of observers
where their subjective experience is recorded.

Taking this work seriously, we may
postulate that the formalism of quantum theory primarily describes the
process of perception and thought. According to such an approach,
the reason for the applicability of quantum theory in the
microscopic phenomena is because the character of our interaction with the 
microscopic world can be viewed in the same way as was done here for the 
case of the perception/thought
process. The stimuli from the external world reach our
consciousness only through the application of the abstract process
of thought and perception \cite{von-neumann}. Therefore, in order
to describe the external world as it is perceived by our 
conscious apparatus, we need to treat it  using the same formulation
used to describe the process of perception and thought. 
Otherwise we would use
two different ways of describing our interaction with 
the external stimuli and this would be
inconsistent.

The experimental activity to
determine the properties of matter at the atomic or subatomic level
can be regarded as an application 
of the process of perception and thought through the elaborate extension
of our sensory apparatus, i.e., using the experimental
instruments which are made using the thought process. The questions we ask, 
when probing the microscopic world, are
based on mentally conceived notions while the ``object of observation''
behaves as ``stimulus'' to the observing device which is an extension
to our sensory apparatus. Furthermore, when we operationally apply our
mental constructs through the observing instruments  to the microscopic world, 
they cause significant effects on the nature of the outcome of the
perception or measurement process. Namely, the nature of the 
questions raised in this case have a determining effect on the
observed and, this clearly forces us to use more explicitly
and more clearly the formalism which describes the process of the operation of
thought and perception. Classical mechanics is just a limit
of quantum theory when the perturbation caused on the observed as
a result of the measurement process is relatively small.

Therefore, in order to describe the operation of 
consciousness and of thought one does not have to explain how the
decoherence is avoided in the brain \cite{tegmark}. 
This becomes an issue only in the case
where one believes that quantum theory is a theory only describing
microscopic particles and, thus, when one tries to explain the brain activity
starting from such a belief it is difficult to understand how 
decoherence effects do not
become important. Here, however, a very different argument is used, namely, 
that  the mental/abstract process of thought
and consciousness is mathematically described by
the formalism of quantum theory as described in Sec.~\ref{formulation}; 
the fact that the formalism of quantum theory is applicable in
the microscopic world {\it follows} from this assertion.
According to this scenario, in order to describe the process of
throught using the formalism of quantum theory, we do not need to 
identify microscopic processes and to show that they survive the effects of
environmental decoherence. It is simply the wrong  basis, the wrong level 
to begin in order to describe the operation of consciousness.

\section{Conclusions}
In summary, the present theory accurately describes: (a) the distribution
of dominance duration in binocular rivalry; (b) the qualitative change
in distribution of dominance duration and the appearance of oscillations
in subjects which were studied under the influence of hallucinogenic
drugs which, as is found, increase the neuron firing rates and decrease
the potential frequency; (c) the marked increase in dominance duration
by periodic removal of the stimulus.

Furthermore, for binocular rivalry experiments where the stimulus
is periodically removed, a distribution of perceptual alterations
as a function of time is predicted and this can be tested in future
experiments.

\section{Acknowledgements}

I wish to thank Henry Stapp for very illuminating discussions. In
addition, thanks are due to Sean Barton for proof-reading the manuscript
and to Stanley Klein and Olivia Carter for their critical comments.

\section{Appendix}

\label{appendixa}

\begin{figure}[htp]
 \vskip 0.2 in 

\begin{centering}
\includegraphics[width=\figwidth]{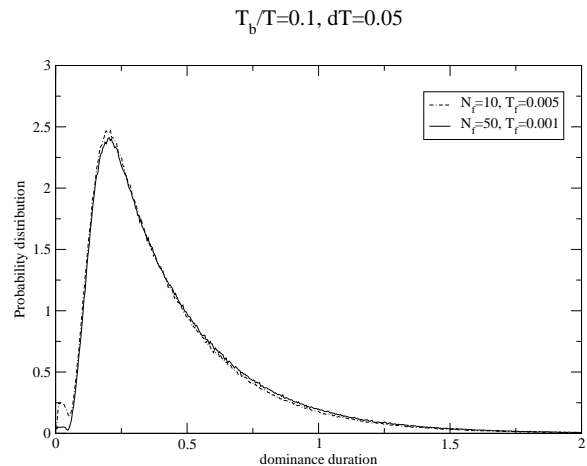} 
\par\end{centering}

\caption{\label{bfcomparison} Comparison of $P(t)$ obtained for different
firing rates but keeping the duration of firing and everything else
constant.}

\vskip 0.2 in 
\end{figure}

\begin{figure}[htp]
 \vskip 0.2 in 

\begin{centering}
\includegraphics[width=\figwidth]{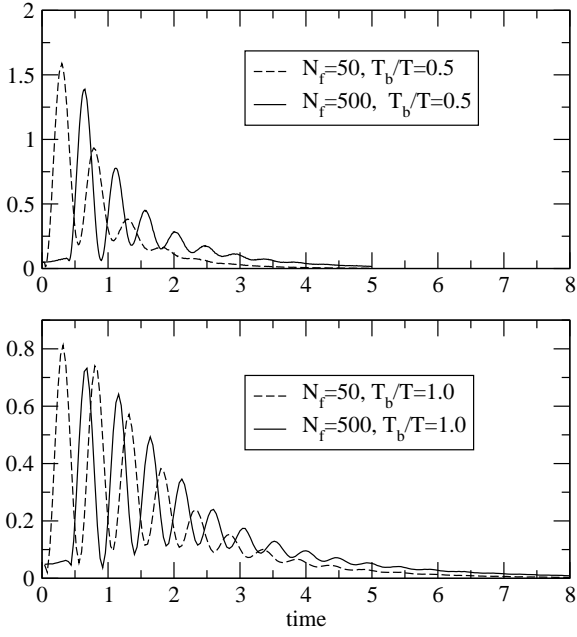} 
\par\end{centering}

\caption{\label{burst} Comparison of $P(t)$ obtained for $T_{f}/T=0.001$
and $N_{f}=50$ and $N_{f}=500$. Top: $T_{b}/T=0.5$. Bottom $T_{b}/T=1.0$.}

\vskip 0.2 in 
\end{figure}

\begin{figure}[htp]
 \vskip 0.2 in 

\begin{centering}
\includegraphics[width=\figwidth]{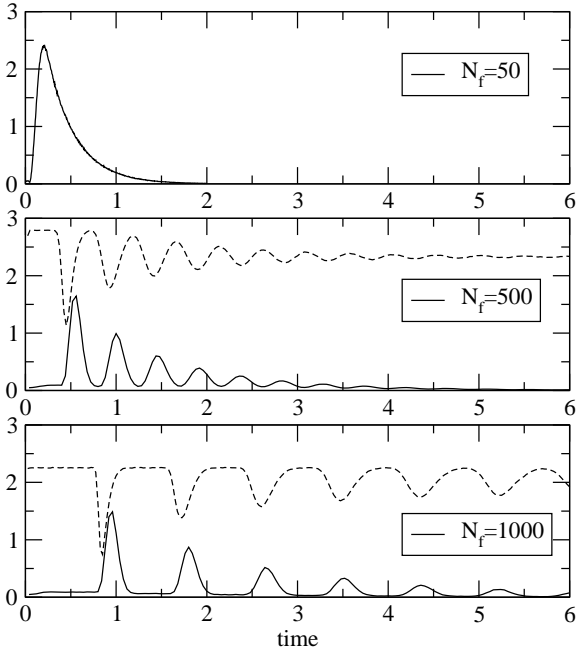} 
\par\end{centering}

\caption{\label{duration} Comparison of $P(t)$ obtained for $T_{b}/T=0.1$
and $T_{f}/T=0.001$. Top: $N_{f}=50$. Middle: $N_{f}=500$. Bottom:
$N_{f}=1000$.}

\vskip 0.2 in 
\end{figure}

\begin{figure}[htp]
 \vskip 0.5 in 

\begin{centering}
\includegraphics[width=\figwidth]{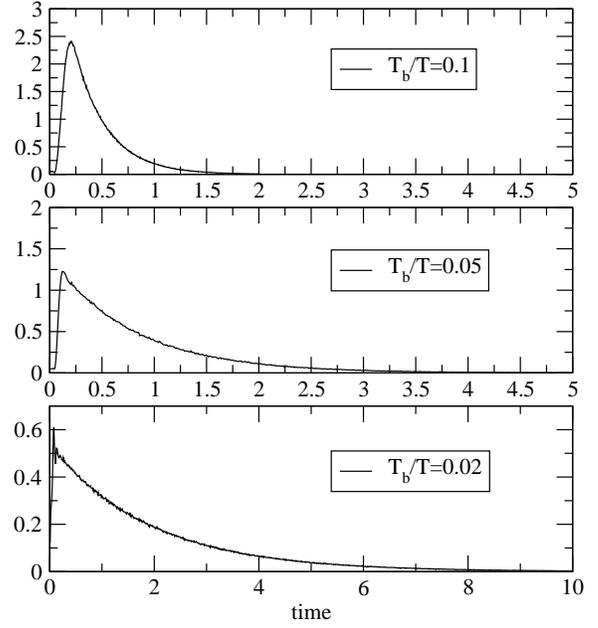} 
\par\end{centering}

\caption{\label{varioussmalleps} Comparison of $P(t)$ obtained for different
small values of $T_{b}/T$ (0.1 (top), 0.05 (center), and 0.02 (bottom)) 
for $T_{f}/T=0.005$ and $N_{f}=10$. The
decay time constant increases by decreasing $T_{b}$. Notice the difference
in the scale of time used in the graphs.}

\vskip 0.2 in 
\end{figure}

\begin{figure}[htp]
 \vskip 0.2 in 

\begin{centering}
\includegraphics[width=\figwidth]{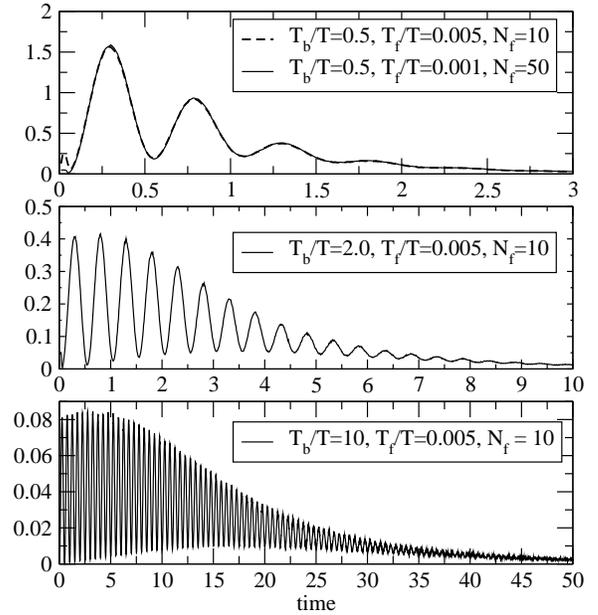} 
\par\end{centering}

\caption{\label{variouseps} Comparison of $P(t)$ obtained for different values
of $T_{b}/T$ (0.5 (top), 2.0 (center), and 10 (bottom)) for 
$T_{f}/T=0.005$ and $N_{f}=10$. Notice the difference
in the scale of time used in the plots.}

\vskip 0.2 in 
\end{figure}

In this appendix we study the role of the parameters which control
the neuron firing rates and the various qualitative different regimes
of the present model.

In Fig.~\ref{bfcomparison} we study the effect of changing the firing
frequency within any given burst. Therefore we keep the other parameters,
namely, the average interval between the end of a burst and the beginning
of the next burst $T_{b}/T=0.1$ and the burst duration $\delta T\equiv N_{f}T_{f}=0.1$,
fixed. Notice that if we change the number of measurements (which
result to neuron firings) within each burst from 10 to 50 the calculated
probability distribution of dominance duration $P(t)$ changes significantly
only for $0<t\le\delta T$. This means that when the 
average interval between measurements
is reduced the probability for perceptual change is significantly
reduced. The reason for that change is the so-called quantum Zeno
effect. This can be easily understood if we consider that the measurements
are done at equally spaced intervals $\delta t=\delta T/N_{f}$. The
expression given for the transition probability gives 
\begin{eqnarray}
P_{N_{f}}(t)=(\cos(\bar{\omega}\delta T/N_{f}))^{2N_{f}}\sin^{2}(\bar{\omega}\delta T/N_{f})
\end{eqnarray}
and for large $N_{f}$ the leading term is \begin{eqnarray}
P_{N_{f}\to\infty}(t)=\Bigl(\frac{{\bar{\omega}\delta T}}{{N_{f}}}\Bigr)^{2}.\label{zeno}\end{eqnarray}

In Fig.~\ref{burst} and Fig.~\ref{duration} we demonstrate the
effect of increasing the burst duration $\delta T$. We need to distinguish
two regimes:

\begin{enumerate}
\item When $\delta T\le T_{b}$. In Fig.~\ref{burst} it is demonstrated
that, by increasing $N_{f}$ from 50 to 500, the initial interval
where $P(t)$ is small (of the order of that given by Eq.~\ref{zeno})
is extended and it is of the order of $\delta T$. The oscillatory
behavior of the function $P(t)$ which is controlled by the period
$T$ is unchanged but with the onset delayed by $\delta T$.
\item When $\delta T>T_{b}$, as in the example of Fig.~\ref{duration}.
In this case an increase of $\delta T$ leads to an oscillatory behavior
with period given approximately by $\delta T$. In Fig.~\ref{duration}
(middle and bottom) the dashed line is the average measurement probability
density. Notice the oscillatory behavior of the measurement probability
density matches the oscillatory behavior of $P(t)$.
\end{enumerate}
In Fig.~\ref{varioussmalleps} the function $P(t)$ calculated for
small values of $T_{b}/T$ (and $\delta T/T=0.05$) is plotted. The
function $P(t)$ for large $t$ decays exponentially with a decay
time constant which increases by decreasing $T_{b}$ as in the quantum
Zeno case.

In Fig.~\ref{variouseps} the function $P(t)$ calculated for larger
values of $T_{b}/T$ (and $\delta T/T=0.05$) is plotted. The function
$P(t)$ is oscillating with period $T$ and with envelop of the oscillation
decays exponentially with a decay time constant which increases by
increasing $T_{b}$.

\bibliographystyle{apalike}

\begin{thebibliography}{}

\bibitem[Achermann and Borbely, 1997]{eeg0}
Achermann, P. and Borbely, A.~A. (1997).
\newblock Low frequency ($< 1$ hz) oscillations in the human sleep
  electroencephalogram.
\newblock {\em Neurosci.}, 81(1):213.

\bibitem[Atmanspacher et~al., 2008]{atmanspachernew}
Atmanspacher, H., Bach, M., Filk, T., Kornmeier, J., and Roemer, H. (2008).
\newblock Cognitive time scales in a necker-zeno model for bistable perception.
\newblock {\em Open Cybernetics and Systemics Journal}, 2:234--251.

\bibitem[Atmanspacher et~al., 2004]{atmanspacher2}
Atmanspacher, H., Filk, T., and Roemer, H. (2004).
\newblock Quantum zeno features of bistable perception.
\newblock {\em Biol. Cybern.}, 90:33.

\bibitem[Atmanspacher et~al., 2002]{atmanspacher1}
Atmanspacher, H., Roemer, H., and Walach, H. (2002).
\newblock Weak quantum theory: complementarity and entanglement in physics and
  beyond.
\newblock {\em Found. Phys.}, 32:379.

\bibitem[Blake and Logothetis, 2002]{visualcompetetion}
Blake, R. and Logothetis, N. (2002).
\newblock Visual competition.
\newblock {\em Nature Rev. Neurosci.}, 3:13.

\bibitem[Bohm, 1979]{bohm}
Bohm, D. (1979).
\newblock {\em Quantum Mechanics}.
\newblock Dover, New York.

\bibitem[Carter and Pettigrew, 2003]{lsd}
Carter, O.~I. and Pettigrew, J.~D. (2003).
\newblock A common oscillator for perceptual rivalries?
\newblock {\em Perception}, 32:295.

\bibitem[Carter et~al., 2007]{carter2}
Carter, O.~L., Hasler, F., Pettigrew, J.~D., Wallis, G.~M., Liu, G.~B., and
  Vollenweider, F.~X. (2007).
\newblock Psilocybin links binocular rivalry switch rate to attention and
  subjective arousal levels in humans.
\newblock {\em Psychopharmacology}, 195:415--424.

\bibitem[Carter et~al., 2005]{carter1}
Carter, O.~L., Pettigrew, J.~D., Hasler, F., Wallis, G.~M., Liu, G.~B., Hell,
  D., and Vollenweider, F.~X. (2005).
\newblock Modulating the rate and rhythmicity of perceptual rivalry
  alternations with the mixed 5-ht2a and 5-ht1a agonist psilocybin.
\newblock {\em Neuropsychopharmacology}, 30:1154.

\bibitem[Chong and Blake, 2005]{chongblake}
Chong, S.~C. and Blake, R. (2005).
\newblock Exogenous and endogenous attention influence initial dominance in
  binocular rivalry.
\newblock {\em J. Vision}, 5(8):1045a.

\bibitem[Chong et~al., 2005]{chong}
Chong, S.~C., Tadin, D., and Blake, R. (2005).
\newblock Endogenous attention prolongs dominance durations in binocular
  rivalry.
\newblock {\em J. Vision}, 5:1004.

\bibitem[Destexhe et~al., 1999]{eegcat}
Destexhe, A., Conteras, D., and Steriade, M. (1999).
\newblock Spatiotemporal analysis of local field potentials and unit discharges
  in cat cerebral cortex during natural wake and sleep states.
\newblock {\em J. Neurosci.}, 19(11):4595.

\bibitem[Engel et~al., 1992]{singer1}
Engel, A.~K., Konig, P., Kreiter, A., Schillen, T.~B., and Singer, W. (1992).
\newblock Temporal coding in the visual cortex: new vistas on integration in
  the nervous system.
\newblock {\em Trends Neurosci.}, 15:218.

\bibitem[Ferri et~al., 2005]{eeg2}
Ferri, R., Bruni, O., Miano, S., Plazzi, G., and Terzano, M.~G. (2005).
\newblock All-night eeg power spectral analysis of the cyclic alternating
  pattern components in young adult subjects.
\newblock {\em Clin. Neurophys.}, 116:2429.

\bibitem[Freeman, 2005]{freeman}
Freeman, A.~W. (2005).
\newblock Multistage model for binocular rivalry.
\newblock {\em J. Neurophysiol.}, 94:4412.

\bibitem[Fries et~al., 2002]{fries}
Fries, P., Schr\"oder, J.-H., Roelfsema, P.~R., Singer, W., and Engel, A.~K.
  (2002).
\newblock Oscillatory neuronal synchronization in primary visual cortex as a
  correlate of stimulus selection.
\newblock {\em J. Neurosci.}, 22:3739.

\bibitem[Giulini et~al., 1996]{Giulini}
Giulini, D., Joos, E., Kieffer, C., Kupsch, J., Stamatescu, I.~O., and Zeh,
  H.~D. (1996).
\newblock {\em Decoherence and the Appearance of a Classical World in Quantum
  Theory}.
\newblock Springer New York.

\bibitem[Gray, 1994]{gray}
Gray, C.~M. (1994).
\newblock Synchronous oscillations in neuronal systems: Mechanisms and
  functions.
\newblock {\em J. Comp. Neurosci.}, 1:11--38.

\bibitem[Gray and Singer, 1989]{singer2}
Gray, C.~M. and Singer, W. (1989).
\newblock Stimulus-specific neuronal oscillations in orientation columns of cat
  visual cortex.
\newblock {\em Proc. Natl. Acad. Sci. USA}, 86:1698.

\bibitem[Hobson, 2005]{eeg1}
Hobson, J.~A. (2005).
\newblock Sleep is of the brain, by the brain and for the brain.
\newblock {\em Nature}, 437:1254.

\bibitem[Joos and Zeh, 1985]{Joos}
Joos, E. and Zeh, H.~D. (1985).
\newblock The emergence of classical properties through interaction with the
  environment.
\newblock {\em Z. Phys.}, B59:223--243.

\bibitem[Jung and Pauli, 2001]{pauli}
Jung, C.~G. and Pauli, W. (2001).
\newblock {\em Atom and the Archetype: Pauli/Jung Letters, 1932-1958}.
\newblock ed. C. A. Meier, (Princeton University Press, Princeton).

\bibitem[Lehky, 1995]{rivalrydata}
Lehky, S.~R. (1995).
\newblock Binocular rivalry is not chaotic.
\newblock {\em Proc, R, Soc, Lond.}, B 259:71.

\bibitem[Leopold and Logothetis, 1999]{leopold}
Leopold, D.~A. and Logothetis, N.~K. (1999).
\newblock Multistable phenomena: changing views in perception.
\newblock {\em Trends of Cogn. Sci.}, 3:254.

\bibitem[Leopold et~al., 2002]{stablerivalry}
Leopold, D.~A., Wilke, M., Mair, A., and Logothetis, N. (2002).
\newblock Stable perception of visually ambiguous patterns.
\newblock {\em Nature Neurosci.}, 5:605.

\bibitem[Levelt, 1968]{levelt}
Levelt, W. J.~M. (1968).
\newblock {\em Psychological Studies on Binocular Rivalry}.
\newblock Mouton, Hague.

\bibitem[London and Bauer, 1983]{london}
London, F. and Bauer, E. (1983).
\newblock {\em Quantum Theory and Measurement}.
\newblock J. A. Wheeler and W. H. Zurek eds., pg 217 (Princeton University
  Press, Princeton).

\bibitem[Lumer et~al., 1998]{fmri2}
Lumer, E.~D., Friston, K.~J., and Rees, G. (1998).
\newblock Neural correlates of perceptual rivalry in the human brain.
\newblock {\em Science}, 280:1930.

\bibitem[Manousakis, 2006]{foundations}
Manousakis, E. (2006).
\newblock Founding quantum theory on the basis of consciousness.
\newblock {\em Found. Phys.}, 36(6):795.

\bibitem[Martinez-Conde et~al., 2000]{monkey}
Martinez-Conde, S., Macknik, S.~L., and Hubel, D.~H. (2000).
\newblock Microsaccadic eye movements and firing of single cells in the striate
  cortex of macaque monkeys.
\newblock {\em Nature Neurosci}, 3:251.

\bibitem[Mavromatos and Nanopoulos, 1998]{nanopoulos}
Mavromatos, N.~E. and Nanopoulos, D.~V. (1998).
\newblock On quantum mechanical aspects of microtubules.
\newblock {\em J. Mod. Phys.}, B 12:517.

\bibitem[Meng and Tong, 2004]{meng}
Meng, M. and Tong, F. (2004).
\newblock Can attention selectively bias bistable perception? differences
  between binocular rivalry and ambiguous figures.
\newblock {\em J. Vision}, 4:539--551.

\bibitem[Mitchell et~al., 2004]{mitchell}
Mitchell, J.~F., Stoner, G.~R., and Reynolds, J.~H. (2004).
\newblock Object-based attention determines dominance in binocular rivalry.
\newblock {\em Nature}, 429:410.

\bibitem[Ooi and He, 1999]{ooi}
Ooi, T.~L. and He, Z.~J. (1999).
\newblock Binocular rivalry and visual awareness: The role of attention.
\newblock {\em Perception}, 28:551.

\bibitem[Orbach et~al., 1966]{orbach}
Orbach, J., Zucker, E., and Olson, R. (1966).
\newblock Reversibility of the necker cube vii. reversal time as a function of
  figure-on and figure-off durations.
\newblock {\em Perceptual and Motor Skills}, 22:615.

\bibitem[Penrose, 1989]{penrose}
Penrose, R. (1989).
\newblock {\em The emperor's new mind}.
\newblock Oxford University Press, NY.

\bibitem[Peterson and Peterson, 1959]{shortmemory}
Peterson, L.~R. and Peterson, M.~J. (1959).
\newblock Shortterm retention of individual verbal items.
\newblock {\em J. Exp. Psychol.}, 58(3):193.

\bibitem[Polonsky et~al., 2000]{fmri3}
Polonsky, A., Blake, R., Braun, J., and Heeger, D.~J. (2000).
\newblock Neuronal activity in human primary visual cortex correlates with
  perception during binocular rivalry.
\newblock {\em Nat. Neurosci.}, 3(11):1153.

\bibitem[P\"oppel, 1997]{poppel}
P\"oppel, E. (1997).
\newblock A hierarchical model of temporal perception.
\newblock {\em Trends Cognit. Sci.}, 1:56--61.

\bibitem[Sams et~al., 1993]{neurosquid}
Sams, M., Hari, R., Rif, J., and Knuutila, J. (1993).
\newblock The human auditory sensory memory trace persists about 10 sec:
  neuromagnetic evidence.
\newblock {\em J. Cogn. Neurosci.}, 5(3):363.

\bibitem[Schroedinger, 1967]{schrodinger}
Schroedinger, E. (1967).
\newblock {\em What is life? and Mind and Matter}.
\newblock Cambridge University Press, Cambridge.

\bibitem[Schwartz et~al., 2005]{stapp2}
Schwartz, J.~M., Stapp, H.~P., and Beauregard, M. (2005).
\newblock Quantum physics in neuroscience and psychology: a neurophysical model
  of mind-brain interaction.
\newblock {\em Phil. Tran. Royal Soc.}, B 360(1458):1306.

\bibitem[Schwartz et~al., 2007]{Schwartz07}
Schwartz, O., Hsu, A., and Dayan, P. (2007).
\newblock Space and time in visual context.
\newblock {\em Nature Rev. Neurosci.}, 8:522.

\bibitem[Singer and Gray, 1995]{singer3}
Singer, W. and Gray, C.~M. (1995).
\newblock Visual feature integration and the temporal correlation hypothesis.
\newblock {\em Annu. Rev. Neurosci.}, 18:555.

\bibitem[Stapp, 1980]{stapp}
Stapp, H.~P. (1980).
\newblock Locality and reality.
\newblock {\em Found. Phys.}, 10:767.

\bibitem[Stapp, 2003]{stapp1}
Stapp, H.~P. (2003).
\newblock {\em Mind, Matter and Quantum Mechanics}.
\newblock Springer-Verlag, Berlin.

\bibitem[Stapp, 2007]{stapp3}
Stapp, H.~P. (2007).
\newblock {\em Mindful Universe: Quantum Mechanics and the Participaring
  Observer}.
\newblock Springer-Verlag, Berlin.

\bibitem[Tegmark, 2000]{tegmark}
Tegmark (2000).
\newblock The importance of quantum decoherence in brain processes.
\newblock {\em Phys. Rev. E.}, 61:4194--4206.

\bibitem[Tong, 2003]{visualawareness}
Tong, F. (2003).
\newblock Primary visual cortex and visual awareness.
\newblock {\em Nature Rev. Neurosci.}, 4:219.

\bibitem[Tong and Engel, 2001]{fmri4}
Tong, F. and Engel, S.~A. (2001).
\newblock Interocular rivalry revealed in the human cortical blind-spot
  representation.
\newblock {\em Nature}, 411:195.

\bibitem[Tong et~al., 2006]{Tong06}
Tong, F., Meng, M., and Blake, R. (2006).
\newblock Neural bases of binocular rivalry.
\newblock {\em Trends Cogn Sci}, 10:512.

\bibitem[Tong et~al., 1998]{fmri}
Tong, F., Nakayama, K., Vaughan, J.~T., and Kanwisher, N. (1998).
\newblock Binocular rivalry and visual awareness in human extrastriate cortex.
\newblock {\em Neuron}, 21:753.

\bibitem[Von-Neumann, 1955]{von-neumann}
Von-Neumann, J. (1955).
\newblock {\em Mathematical Foundations of Quantum Mechanics, Chap. VI, pg.
  417}.
\newblock Princeton University Press, Princeton.

\bibitem[Wigner, 1983]{wigner}
Wigner, E.~P. (1983).
\newblock {\em Quantum Theory and Measurement}.
\newblock J. A. Wheeler and W. H. Zurek eds., pg 260 and 325 (Princeton
  University Press, Princeton).

\bibitem[Zeh, 1970]{Zeh}
Zeh, H.~D. (1970).
\newblock Toward a quantum theory of observation.
\newblock {\em Found. Phys.}, 1:69--76.

\end{thebibliography}

\end{document}